\documentclass[a4paper, amsfonts, amssymb, amsmath, reprint, showkeys, nofootinbib, twoside, superscriptaddress]{revtex4-1}
\usepackage[english]{babel}
\usepackage[utf8]{inputenc}
\usepackage{siunitx}
\usepackage[colorinlistoftodos, color=green!40, prependcaption]{todonotes}
\usepackage{amsthm}
\usepackage{mathtools}
\usepackage{physics}
\usepackage{xcolor}
\usepackage{graphicx}
\usepackage[left=23mm,right=13mm,top=35mm,columnsep=15pt]{geometry} 
\usepackage{adjustbox}
\usepackage{placeins}
\usepackage[T1]{fontenc}
\usepackage{lipsum}
\usepackage{csquotes}
\usepackage{caption}
\usepackage{wrapfig}
\usepackage{gensymb}

\usepackage{comment}
\usepackage[pdftex, pdftitle={Article}, pdfauthor={Author}]{hyperref} 
\begin{document}
\title{Evaluation of quantum key distribution systems against injection-locking attacks}

\author{Jerome Wiesemann}
\thanks{These authors contributed equally.}
\affiliation{Fraunhofer Institute for Telecommunications, Heinrich Hertz Institute, HHI, 10587 Berlin, Germany}

\author{Fadri Grünenfelder}
\thanks{These authors contributed equally.}
\affiliation{Vigo Quantum Communication Center, University of Vigo, Vigo E-36310, Spain}
\affiliation{Escuela de Ingeniería de Telecomunicación, Department of Signal Theory and Communications, University of Vigo, Vigo E-36310, Spain}
\affiliation{AtlanTTic Research Center, University of Vigo, Vigo E-36310, Spain}

\author{Ana Blázquez}
\affiliation{Vigo Quantum Communication Center, University of Vigo, Vigo E-36310, Spain}
\affiliation{Escuela de Ingeniería de Telecomunicación, Department of Signal Theory and Communications, University of Vigo, Vigo E-36310, Spain}
\affiliation{AtlanTTic Research Center, University of Vigo, Vigo E-36310, Spain}

\author{Nino Walenta}
\affiliation{Fraunhofer Institute for Telecommunications, Heinrich Hertz Institute, HHI, 10587 Berlin, Germany}

\author{Davide Rusca}
\affiliation{Vigo Quantum Communication Center, University of Vigo, Vigo E-36310, Spain}
\affiliation{Escuela de Ingeniería de Telecomunicación, Department of Signal Theory and Communications, University of Vigo, Vigo E-36310, Spain}
\affiliation{AtlanTTic Research Center, University of Vigo, Vigo E-36310, Spain}

\date{\today} 

\begin{abstract}
 While ideal quantum key distribution (QKD) systems are well-understood, practical implementations face various vulnerabilities, such as side-channel attacks resulting from device imperfections. Current security proofs for decoy-state BB84 protocols either assume uniform phase randomization of Alice's signals, which is compromised by practical limitations and attacks like injection locking, or rely on a (partially) characterized phase distribution. This work presents an experimental method to characterize the phase de-randomization from injection locking using a heterodyne detection setup, providing a lower bound on the degree of isolation required to protect QKD transmitters against injection-locking attacks. The methods presented are source-agnostic and can be used to evaluate general QKD systems against injection-locking attacks.
\end{abstract}

\maketitle

\section{Introduction} \label{sec:introduction}
\noindent Quantum key distribution (QKD) is a method of establishing information-theoretically secure keys between two parties, commonly known as Alice and Bob \cite{Bennett84, Gisin02, Scarani09}. Unlike classical methods that rely on computational assumptions about a potential adversary Eve, QKD provides a solution where Eve is only limited by the laws of quantum physics. While the security of ideal QKD systems is rather well-understood \cite{Renner05, Tomamichel12, Lim14, Tomamichel17, Rusca18}, security proofs make various model assumptions about the physical implementation \cite{Zapatero23, Curras24}. Any discrepancy between the model and the implementation, for example due to device imperfections, may lead to \textit{side-channel attacks}, where Eve gathers additional information about the key which is not modeled in the proof \cite{Sajeed21, Makarov2023, BSI24}. Usually, the model must then be adjusted to account for given device imperfections and security proven for this new model. In the last two decades, much effort has been devoted to closing the gap between the implementation and the model, e.g. by increasing the complexity of the theoretical models \cite{Curras24, Tupkary24, Sixto2024}.

As a well-known example, earlier analyses assumed that Alice generates single photons \cite{Gottesman04, Tomamichel12}. Owing to their greater practicality, weak coherent states have become a favored mean to generate Alice's signals as part of the \textit{decoy-state BB84 protocol} \cite{Hwang03, wang2005, Lo05}, which is arguably the most widespread protocol today due to its high practicality and key rates. A common assumption of security proofs for decoy-state QKD is that the phases of Alice's signals are uniformly random \cite{Lim14, Rusca18, Tupkary24}. Security can still be proven without uniformly distributed phases but the performance is drastically affected with current security proofs \cite{Sixto23, Nahar23, Curras24} and requiring a (partial) characterization of the degree of phase randomization. 

The \textit{injection-locking attack} specifically aims at attacking the phase randomization process by injecting light into Alice's setup in order to lock the phase of her light source.  We note that even without this attack, perfect uniform distribution is experimentally not feasible and the degree of phase randomization must be experimentally characterized. Laser-seeding attacks have been experimentally demonstrated using phase information \cite{Tang13} and frequency information \cite{Xiao-Ling20}. The effect of a laser-seeding attacks on the intensity has been studied experimentally and theoretically in \cite{Huang19}. Injection locking has also been used for a photon-number splitting attack \cite{Xiao-Xu22}. A recent work describes the effect of injection locking on phase de-randomization \cite{Lovic2023} but their analysis focuses on modeling the laser behavior and is thus not generally applicable without specific assumptions about the laser behavior. Until now, analyses of injection locking focused on experimental demonstrations but do not provide a rigorous method to evaluate QKD systems against such attack.

Our work specifically aims at proposing an approach to experimentally characterize the degree of phase randomization of QKD systems under injection locking using a metric similar to recent theoretical analyses of imperfection phase randomization \cite{Nahar23, Curras2024}. We present a robust method to evaluate QKD transmitters against phase de-randomization from injection locking. The main result is a characterization yielding a minimum degree of isolation required to protect QKD transmitters against injection-locking attacks. The experimental methods presented are source-agnostic and can be used to evaluate QKD systems against injection-locking attacks, e.g. in evaluation laboratories through black-box testing as part of a certification process.  While we demonstrate the methods using a DFB laser in this work, the device under test (DUT) can be substituted, for instance, with a QKD transmitter. 

In Sec.~\ref{sec:background}, we begin by discussing the role of phase randomization in QKD and introduce the $q_\mathrm{rel}$-parameter as a metric to quantify the degree of phase de-randomization under an injection-locking attack. We also describe the injection-locking attack and discuss how it affects the security of QKD systems. We then present the methods used to determine the relative phase distribution of the DUT using a heterodyne detection setup in Sec.~\ref{sec:characterization} and discuss the effect of polarization on the degree of injection locking. Finally, the main result is a characterization of the $q_\mathrm{rel}$-parameter as a function of the injected optical power, which yields a lower bound on the degree of optical isolation required to protect QKD systems against injection-locking attacks.

\section{Background} \label{sec:background}
\subsection{Phase randomization in QKD} \label{sec:phase_rand_in_QKD}
\noindent In most of today's security proofs for the decoy-state BB84 protocol, Alice is assumed to send phase-randomized weak coherent states \cite{Lo2007, Lim14, Rusca18, Wiesemann24, Tupkary24}. Uniform phase randomization is a core assumption of these proofs. Indeed, a state with intensity $\mu$ sent by a phase-randomized coherent source can be represented by the density matrix 
\begin{equation}
    \rho_\mu = \int_0^{2\pi} d\theta f(\theta) \ket{\sqrt{\mu}e^{i\theta}}\bra{\sqrt{\mu}e^{i\theta}}, 
    \label{eq:density_matrix_coherent}
\end{equation}
where $\ket{\sqrt{\mu}e^{i\theta}}$ are coherent states with mean photon number $\mu$ and phase $\theta$ \cite{Glauber1963}, and $f$ is the probability density function (PDF) representing the probability of sending a coherent state with a given phase. In the case of uniformly distributed phases, $f(\theta) = 1 / (2\pi)$, and the equation above can be rewritten as a coherent superposition of Fock states $\ket{n}$ (see App.~\ref{ap:diagonal_form})
\begin{equation}
    \rho_\mu = \sum_{n=0}^{\infty} e^{-\mu} \frac{\mu^n}{n!}\ket{n}\bra{n}\,,
    \label{eq:density_matrix_fock}
\end{equation}
i.e. it is diagonal in the Fock basis. This is used in security proofs to reduce the problem to a discussion of the statistics of vacuum, single-photon and multi-photon events \cite{Lim14, Rusca18, Wiesemann24, Tupkary24}. In the case of imperfect phase randomization, e.g. due to device imperfections or active attacks such as injection locking, Eq.~\eqref{eq:density_matrix_fock} does not hold and the standard decoy-state BB84 method cannot be applied. It is still possible to prove security without uniform phase randomization but current security proofs \cite{Nahar23, Sixto23, Curras2024} require a characterization of either the full PDF $f(\theta)$ or a metric $q$ describing the degree of phase randomization, which we introduce in Sec.~\ref{sec:quantifying_phase_randomization}. We will use a metric closely related to the $q$-parameter to quantify the degree of phase de-randomization induced by Eve from an injection-locking attack. This metric provides a practical means for evaluating the security of QKD systems against these attacks.

\subsection{Injection-locking attack}
\noindent There exists two commonly used methods to practically implement phase randomization in QKD: incorporating a phase modulator or using a gain-switched laser. The drawback of the former approach is that it requires an additional active component, a phase modulator, combined with a high-quality entropy source for selecting the random phase, and that the phase choices are discrete, which negatively affects the performance compared to uniform phase randomization \cite{Cao15, Sixto23}. A popular alternative are gain-switched lasers, which inherently produce phase-randomized pulses. Indeed, each gain-switching cycle, the pulse is generated from spontaneous emission and amplification of photons with a random phase. This however requires that spontaneous emission initiates the buildup, which implies that no residual photons remain in the cavity between pulses. If, however, residual photons remain in the cavity, they can initiate the process through stimulated emission, leading to phase correlations between these residual photons and the newly generated pulse.

The injection-locking attack precisely aims at making use of this property. In fact, Eve injects light with a known phase into Alice's laser cavity such that, following the discussion above, the phase of the generated pulses is correlated with that of the injected light. This alters the PDF describing Alice's phase randomization such that, generally, Eq.~\eqref{eq:density_matrix_coherent} does not simplify to Eq.~\eqref{eq:density_matrix_fock} anymore.

To the best of our knowledge, an experimental method to rigorously characterize the influence of Eve on the phase randomization process, by means of injection locking, has not yet been developed and is thus the aim of this work. The methods presented can be used to evaluate QKD systems against injection-locking attacks, e.g. through means of black-box testing. In the following, we use a gain-switched laser as it is vulnerable to these attacks but note that our methods can also be used to characterize other light sources.

\subsection{Quantifying phase randomization} \label{sec:quantifying_phase_randomization}
\noindent We quantify the influence of Eve on the phase randomization by choosing a metric closely related to the one introduced in two recent theoretical analyses of imperfect phase randomization \cite{Nahar23, Curras2024}. To simplify the discussion, we assume that no phase correlations are present in the laser cavity without an injection-locking attack, such that Alice's signals are independently and identically prepared. We note that this assumption is reasonable, as phase correlations at a $5\,$GHz repetition rate are already small \cite{Grunenfelder20}, and we are operating at a repetition rate that is two orders of magnitude lower in this work.
The degree of phase randomization of a sequence of $N$ phases is described by the $q$-parameter defined as \cite{Curras2024}
\begin{equation}
    f\big(\theta^{(n)} |\theta^{(n-1)}...\theta^{(1)}\big)\geq \frac{q}{2\pi}\,,
    \label{eq:q_param_def_guillermo}
\end{equation}
for all $\theta^{(n)} \in [0,2\pi)$, where $n\in\{1, ..., N\}$ and $0<q\leq 1$. In other words, the $q$-parameter is a lower bound on the probability of generating a pulse with any given phase. For uniformly distributed phases, $q=1$, while for imperfect phase randomization $q<1$ and $q$ approaches zero if, for example, the phase distribution is strongly localized. 

Hence, the $q$-parameter describes the degree of phase randomization of Alice's pulses. However, the methods we present don't allow for a direct measurement of the phase $\theta^{(n)}$ but rather the phase $\Delta \theta^{(n)}$ relative to a reference phase, cf. Sec.~\ref{sec:pulse_phase}. Hence, we quantify the degree of phase randomization, relative to a reference phase, by defining a metric similar to the $q$-parameter, which we call \textit{relative $q$-parameter} and denote $q_\mathrm{rel}$, defined as the lower bound
\begin{equation}
    f\big(\Delta\theta^{(n)} | \Delta\theta^{(n-1)} ... \Delta\theta^{(1)}\big) \geq \frac{q_\mathrm{rel}}{2\pi}\,,
    \label{eq:q_rel_param_def}
\end{equation}
for all $\theta^{(n)} \in [0,2\pi)$, where $n\in\{1, ..., N\}$ and $0<q_\mathrm{rel}\leq 1$, analogously to Eq.~\eqref{eq:q_param_def_guillermo}.  We use this metric to quantify the influence of Eve on the phase randomization when performing an injection-locking attack. While it does not directly correspond to the $q$-parameter, it is closely related as it quantifies the degree of phase randomization relative to Eve's reference phase. In fact, $q = q_\mathrm{rel}$ in the case of an ideal attack. We discuss this more thoroughly in Sec.~\ref{sec:pulse_phase}.

As discussed in the previous section, if Eve injects light into the laser cavity, she can modify the PDF describing the phase randomization of Alice's signals. In theory, Eve can introduce correlations between any two pulses by changing the degree of injection locking based on previous measurement outcomes (e.g. by adjusting the optical power or polarization state of her laser). Nevertheless, if the smallest $q_\mathrm{rel}$ she can induce from an injection-locking attack is known, then Eq.~\eqref{eq:q_rel_param_def} holds even if Eve arbitrarily varies the degree of injection locking from pulse to pulse. Consequently, we can ignore correlations in our analysis and focus on determining the smallest $q_\mathrm{rel}$ Eve can induce with an injection-locking attack. To simplify the notation, we define
\begin{equation}
    f\big(\Delta\theta^{(n)}\big) \coloneqq f\big(\Delta\theta^{(n)} | \Delta\theta^{(n-1)} ... \Delta\theta^{(1)}\big)
\end{equation}
in the following. In the next section, we show how to determine the phase distribution of Alice's pulses under an injection-locking attack, relative to a reference phase, and then compute the $q_\mathrm{rel}$-parameter.
\section{Phase randomization characterization} \label{sec:characterization}

\subsection{Setup}
\label{sec:setup}

\begin{figure*}[t]
    \begin{minipage}{\textwidth}
        \centering
        \includegraphics[width=0.9\textwidth]{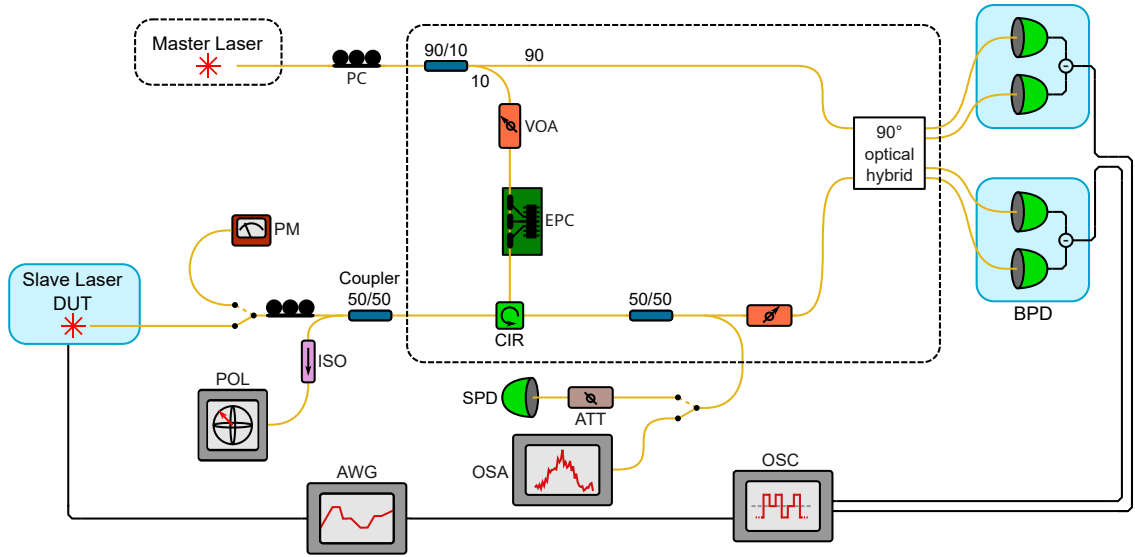}
         \caption{Setup used to characterize the phase distribution under an injection-locking attack. The dashed boxes indicate temperature-stabilized environments. PC: polarization controller, VOA: variable optical attenuator, EPC: electronic polarization controller, CIR: circulator, BPD: balanced photodiodes, SPD: single photon detector, ATT: fixed optical attenuator, OSA: optical spectrum analyzer, OSC: oscilloscope, AWG: arbitrary waveform generator: POL: polarizer, ISO: isolator, PM: powermeter, DFB: distributed feedback laser, DUT: device under test.}
        \label{fig:characterization_setup}
    \end{minipage}
\end{figure*}
\noindent The goal is to quantify the degree of phase de-randomization Eve can induce with an injection-locking attack. In order to achieve this, a heterodyne detection setup with a continuous DFB master laser (ML), corresponding to Eve, and a gain-switched DFB slave laser (SL), corresponding to Alice, is used. The setup is illustrated in Fig.~\ref{fig:characterization_setup} and the equipment listed in App.~\ref{ap:list_equipment}. The ML injects light into the SL and the phase correlations are measured using an interferometer and two homodyne detectors. By increasing the isolation between the ML and SL, using a variable optical attenuator (VOA), we can simulate Alice's components and determine the degree of phase de-randomization Eve can induce with her ML as a function of the optical power reaching Alice's laser, cf. Sec.~\ref{sec:max_degree_il}. We note that, in this work, the DUT is a gain-switched DFB laser but it can be replaced by a QKD transmitter to perform black-box testing, cf. Sec.~\ref{sec:max_degree_il}. 

\begin{figure*}[t]
    \begin{minipage}{\textwidth}
        \centering
        \includegraphics{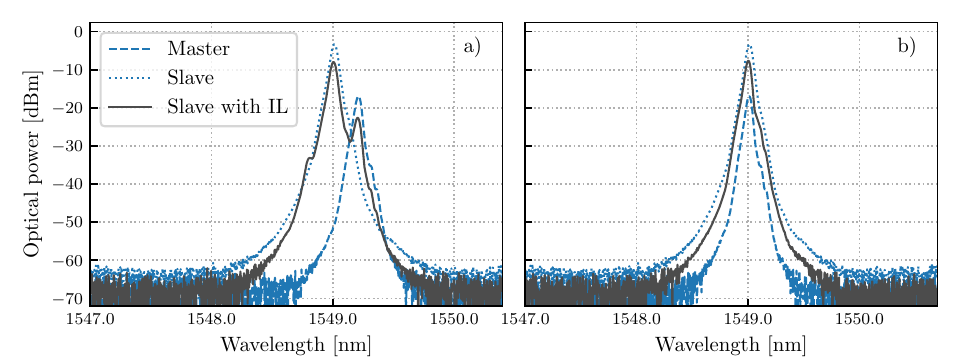} 
        \caption{Spectra of the master laser (ML) and slave laser (SL) as well the SL spectrum under light injection for a) slight spectrum offset ($\vartheta_\text{ML}=20\degree$C) and b) matching spectra ($\vartheta_\text{ML}=17.9\degree$C). The ML central wavelength can be controlled by adjusting the temperature of the laser $\vartheta_\text{ML}$. Note that a $5$\,dB attenuator is placed in front of the OSA  when acquiring the SL spectrum with light injection. }
        \label{fig:laser_spectra}
    \end{minipage}
\end{figure*}

We use master and slave lasers with similar spectra, as depicted in Fig.~\ref{fig:laser_spectra}. The matching of the spectra may significantly affect the degree of injection locking. We thus note an important property of this method, namely that the characterization may only be as good as the ML simulates Eve's best possible attack. In fact, the methods presented in this work provide a lower bound on the degree of injection locking Eve can induce. We remark that the dependency of the experimental characterization on the ML is due to injection locking strongly depending on the specific properties of the devices involved and that there may not exist a single light source which replicates an ideal attack for all DUTs. An alternative approach would involve adopting a different characterization method altogether, though this requires further experimental and theoretical investigations. Nevertheless, we note that this dependency on the specific equipment used is an inherent limitation of an evaluation process conducted as black-box testing.

The SL spectrum under light injection is depicted in Fig.~\ref{fig:laser_spectra} for matching ML and SL spectra and with slight offset. The central wavelength of the ML can be controlled by adjusting the temperature of the laser. The spectra are acquired using an optical spectrum analyzer (OSA) placed at the SL input and output to determine the ML and SL spectra, respectively. The acquisition was performed with a resolution of \SI{0.05}{\nm}.

The ML injects continuous light, with a bandwidth of 1 MHz according to the manufacturer, into the SL which is in turn gain-switched using an arbitrary waveform generator (AWG). The SL is modulated with a $40$\,MHz rectangular signal with 50\% duty cycle. As discussed in Sec.~\ref{sec:phase_rand_in_QKD}, under normal conditions, the gain-switching results in a random phase between each pulse, following the PDF $f\big(\theta^{(n)}\big)$. Now, due to light injection from the ML, the phase of the SL is correlated with that of the ML resulting in a different PDF. We cannot directly determine the phase distribution, as the phase measurement is inherently relative to a references phase. However, we can use the $q_\mathrm{rel}$-parameter, introduced in Sec.~\ref{sec:quantifying_phase_randomization}, as a metric to quantify the phase distribution of the SL, relative to the ML. This motivates why the characterization depends on the ability for the ML to simulate Eve's best possible attack. 

The SL phase distribution, relative to the ML phase, is determined by interfering the ML signal with the SL signal using an interferometer. In the following, the ML will be refereed to as the local oscillator and the SL as the signal. The interferometer is maintained in a thermally controlled environment to minimize thermal fluctuations and the resulting phase fluctuations due to the long interferometer path lengths. We require that the phase fluctuations introduced by the interferometer happen on timescales longer than the acquisition time. In other words, phase fluctuations must be negligible during the time period over which the signal is recorded to produce the histogram discussed below, i.e. Fig.~\ref{fig:histogram}. Otherwise the resulting phase distribution may be more uniformly distributed than expected\footnote{Alternatively, the phase fluctuations may be characterized and the resulting phase distribution adjusted accordingly.}. We found that the change in phase induced by the interferometer during the time of one measurement ($200\,\mu$s) is less than $5\cdot 10^{-4}$\,rad and thus negligible. This was found by replacing the SL with a Faraday mirror and thus interfere the ML with itself.

An exemplary homodyne detection signal, measured on the oscilloscope, is depicted in Fig.~\ref{fig:exemplary_pulse} with injection locking. In the following section, we  determine the relative phase of the SL pulses from the homodyne detection signal.

\subsection{Determining the relative phase of the pulses}
\label{sec:pulse_phase}
\noindent In the following, we denote $T_\mathrm{S}$ the SL period and introduce $n$ to denote the $n$-th pulse. Let $\tau_n = t - nT_\mathrm{S}$ describe the time coordinate relative to the $n$-th pulse. In the following, for any function with subscript $n$, the time dependency is relative to the $n$-th pulse. We write the electric field of the $n$-th pulse of the SL at a fixed point in space as
\begin{equation}
    \mathbf{E}^{(n)}_\mathrm{S}(\tau_n) = e^{i\theta^{(n)}(\tau_n)}\mathbf{A}_\mathrm{S}^{(n)}(\tau_n)\,,
\end{equation}
where $\theta^{(n)}(\tau_n)$ describes the phase of the optical pulse, which is generally random, and
\begin{equation}
    \mathbf{A}_\mathrm{S}^{(n)}(\tau_n) = \left(\mathbf{A}_\mathrm{S}^{(n)}(\tau_n)\right)^*\,.
\end{equation}
Notice that this expression is general and we don't make any assumption about the shape of the pulse nor its spectrum. We also write the LO signal as a generic field
\begin{equation}
    \mathbf{E}_\mathrm{LO}^{(n)}(\tau_n) = e^{i\theta_\mathrm{LO}^{(n)}(\tau_n)}\mathbf{A}_\mathrm{LO}^{(n)}(\tau_n)\,,
\end{equation}
where $\mathbf{A}_\mathrm{LO}^{(n)}(\tau_n) = \left(\mathbf{A}_\mathrm{LO}^{(n)}(\tau_n)\right)^*$. 
Assuming perfectly balanced photodetectors, the homodyne detection signal of the $n$-th pulse is then given by (see App.~\ref{ap:homodyne_detection_signal})
\begin{equation}
    I_0^{(n)} (\tau_n) =  A(\tau_n) \cos\left(\theta^{(n)}(\tau_n) - \theta_\mathrm{LO}^{(n)}(\tau_n) \right)\,,
    \label{eq:I_0_def}
\end{equation}
where $A(\tau_n) \coloneqq 2 \mathbf{A}_\mathrm{S}^{(n)}(\tau_n) \cdot \mathbf{A}_\mathrm{LO}^{(n)}(\tau_n)$. Similarly, the $90\degree$ phase shifted homodyne detection signal is
\begin{equation}
    I_{\pi/2}^{(n)} (\tau_n) =  A(\tau_n) \sin\left(\theta^{(n)}(\tau_n) - \theta_\mathrm{LO}^{(n)}(\tau_n) \right)\,.
    \label{eq:I_pi_2_def}
\end{equation}
Thus, the phase difference between the SL and ML for the $n$-th pulse at time $\tau_n$ is given by
\begin{align}
    \Delta \theta^{(n)}(\tau_n) = \arctan2\left(I_0^{(n)}(\tau_n), I_{\pi/2}^{(n)}(\tau_n)\right)\,,
    \label{eq:expression_theta_mod}
\end{align}
where $\arctan2$ denotes the 2-argument arctangent and we define $\Delta \theta^{(n)}(\tau_n) \coloneqq \theta^{(n)}(\tau_n) - \theta_\mathrm{LO}^{(n)}(\tau_n)$ to simplify the notation.
Intuitively, this phase difference describes the phase of Alice's pulses relative to Eve's reference phase. We see that a heterodyne detection setup yields the relative phase, hence why we introduced the $q_\mathrm{rel}$-parameter in Sec.~\eqref{sec:quantifying_phase_randomization}. We also observe that the $q$-parameter and $q_\mathrm{rel}$-parameter are closely related as, in the case of an ideal attack, $\theta^{(n)}_\mathrm{LO}(\tau_n) = \theta^{(m)}_\mathrm{LO}(\tau_m)$ for all $m, n$, such that $q_\mathrm{rel} = q$ as $f\big(\Delta \theta^{(n)}\big) \equiv f\big(\theta^{(n)}\big)$ up to a constant offset. By definition, we have $q_\mathrm{rel} \geq q$.

We note that the description above allows for general time-dependent phase distributions. Hence, in the following, we add a time-dependency to the PDF $f\large(\Delta \theta^{(n)}, \tau_n\large)$ to describe the relative phase distribution for each point $\tau_n$ on the pulse. Similarly, we denote $q_\mathrm{rel}(\tau_n)$ the relative $q$-parameter corresponding to $f\large(\Delta \theta^{(n)}, \tau_n\large)$, following Eq.~\eqref{eq:q_rel_param_def}. This dependency on $\tau_n$ of the relative phase randomization stems from the fact that the ML is not ideal and that the phase randomization process (and phase shift) of the SL may vary over the duration of the pulse. The latter observation is an effect which, to the best of our knowledge, is not currently incorporated in security proofs which assume a modulation of a constant phase shift over the pulse duration \cite{Sixto23, Nahar23, Curras2024}.

\begin{figure}
    \centering
    \includegraphics{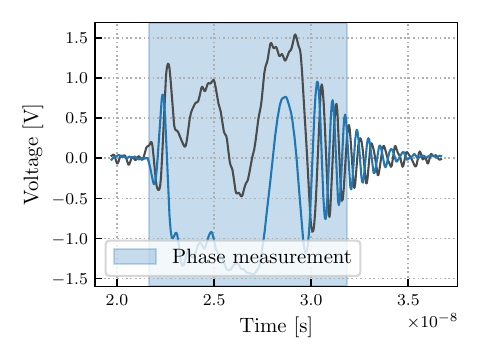}
    \caption{Exemplary oscilloscope homodyne detection signal with $0\degree$ (blue) and $90\degree$ (black) phase shift with $-42.3$\,dBm injection locking. The blue region represents the window used to determine the phase of the pulse.}
    \label{fig:exemplary_pulse}
\end{figure}
In the following, for each measurement, we choose an acquisition time of $200$\,$\mu$s on the oscilloscope, which corresponds to $N = 8000$ pulses. As discussed before, any phase fluctuation, e.g. resulting from temperature fluctuations, must be negligible during the acquisition time. On the other hand, a longer acquisition time implies more samples and thus a better confidence interval on the $q_\mathrm{rel}$-parameter. The oscilloscope has a finite sampling rate (of $50$\,Gsps in our case) and we determine the relative phase $\Delta\theta^{(n)}(\tau_n)$ at each measurement point $\tau_n$, following Eq.~\eqref{eq:expression_theta_mod}. We assume that fluctuations of the relative phase are negligible within one sampling period. The minimum sampling rate required for the characterization thus depends on the pulse width and the desired granularity. 

\begin{figure}[h]
    \centering
    \includegraphics{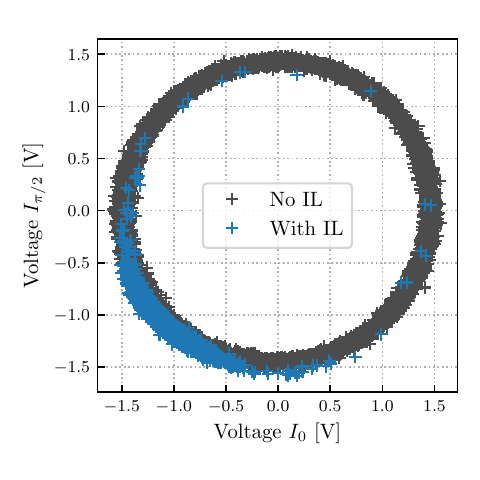}
    \caption{Homodyne detection signals $I^{(n)}_0(\tau_n)$ and $I^{(n)}_{\pi/2}(\tau_n)$, of the $N = 8000$ acquired slave laser pulses, plotted in phase space for $\tau_n = 6.56$\,ns with $-42.3$\,dBm and without injection locking.}
    \label{fig:phase_space_circle}
\end{figure}

We choose a time window which is centered in the pulse to determine the phase, cf. Fig.~\ref{fig:exemplary_pulse}, as the intensity of the homodyne detection signal is too low on the edges of the pulse (and outside of the pulse) for a reliable characterization of the $q_\mathrm{rel}$-parameter. This is an experimental limitation and using, for example, a faster oscilloscope may allow for a characterization over a broader time window. See App.~\ref{app:time_window_argumentation} for a more thorough discussion on how the window is chosen. For each sample at time $\tau_n$ in this window, we determine the corresponding relative phase $\Delta\theta^{(n)}(\tau_n)$ following Eq.~\eqref{eq:expression_theta_mod}. This can be visualized as follows. For each pulse, the homodyne detection signal acquired by the oscilloscope at a given time $\tau_n$ is plotted in phase space where the $x$-coordinate is given by $I_0^{(n)}(\tau_n)$ and the $y$-coordinate is given by $I_{\pi/2}^{(n)}(\tau_n)$, as depicted in Fig.~\ref{fig:phase_space_circle}, for $-42.3$\,dBm light injection and without injection locking. We acquire the measurements without injection locking by disconnecting the fiber between the ML and SL. Following Eq.~\eqref{eq:expression_theta_mod}, the instantaneous relative phase $\Delta \theta^{(n)}(\tau_n)$ between the $n$-th pulse of the SL and the ML is then given by the angle between the vectors $(x, y)^\text{T}$ and $(1, 0)^\text{T}$.

For all $\tau_n$ in the window described above, we then plot the histogram of the phases for a given injected optical power, as depicted in Fig.~\ref{fig:histogram} for $\tau_n = 6.56$\,ns with and without light injection. We clearly observe that the relative phase of the generated optical pulses is localized under an injection-locking attack while almost uniformly distributed without injection locking.

In this section, we have seen how to determine the relative phase of the SL pulses from the heterodyne detection signal. In the next section, we quantify the degree of phase de-randomization from injection locking by determining the $q_\mathrm{rel}$-parameter introduced in Sec.~\ref{sec:quantifying_phase_randomization} from the histogram depicted in Fig.~\ref{fig:histogram}.

\subsection{Determining the degree of phase de-randomization from injection locking}
\label{sec:determine_q_param}
\begin{figure}[h]
    \centering
    \includegraphics{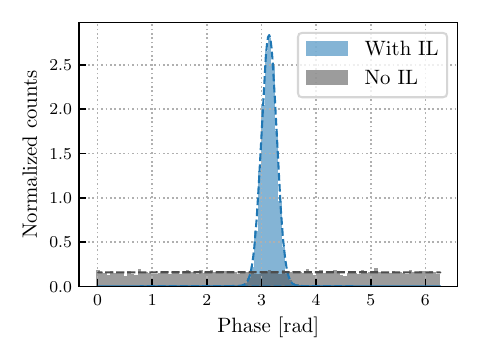}
    \caption{Histogram of the phases for the $N = 8000$ slave laser pulses, determined from Fig.~\ref{fig:phase_space_circle}, for $\tau_n = 6.56$\,ns with $-42.3$\,dBm and without injection locking. }
    \label{fig:histogram}
\end{figure}

\noindent The $q_\mathrm{rel}$-parameter introduced in Sec.~\ref{sec:quantifying_phase_randomization} describes the phase randomization of the SL, relative to the ML. The next step is to determine the $q_\mathrm{rel}$-parameter from the histogram depicted in Fig.~\ref{fig:histogram}, for each $\tau_n$. The histogram yields information about the probability of finding a phase in a certain interval, which is given by the bin width. However, the $q_\mathrm{rel}$-parameter is defined in terms of the PDF $f\big(\Delta\theta^{(n)}, \tau_n\big)$. As such, we choose a suitable model to fit the histogram with. The model used is a wrapped Voigt profile
\begin{align}\label{eq:pdf_model}
f_w\left(\phi,\mu,\sigma,\gamma\right)=\sum_{k=-\infty}^{\infty}V\left(\phi+2\pi k,\mu,\sigma,\gamma\right),
\end{align}
where $V(\phi, \mu, \sigma, \gamma)$ is the uncentered Voigt profile, $\mu$ is the median of the uncentered profile and $\sigma$ is the variance of the normal distribution and $\gamma$ is the scaling parameter of the Cauchy distribution. We fit the data using nonlinear regression and truncate the sum at $k=\pm 10$. We emphasize that the model chosen is part of the assumptions underlying the characterization. Another suitable model can be chosen to fit the histogram.

\noindent We recall that this method does not assume the phase shift to be constant over the pulse duration. We define the minimum $q_\mathrm{rel}$-parameter as
\begin{equation}
    q_\mathrm{rel}^{\mathrm{min}} = \min_{\tau_n} q_\mathrm{rel}(\tau_n)\, ,
    \label{eq:def_q_rel_min}
\end{equation}
where we recall that $q_\mathrm{rel}(\tau_n)$ is defined as the lower bound
\begin{equation}
    f(\Delta\theta^{(n)}, \tau_n) \geq \frac{q_\mathrm{rel}(\tau_n)}{2\pi}\,.
\end{equation}
We note that with enough samples it may be possible to apply suitable concentration inequalities to directly determine the minimum of the PDF corresponding to the phase histogram, cf. Fig.~\ref{fig:histogram}, without making any model assumption. By choosing a small enough bin width, one can assume that the phase distribution remains approximately constant within each bin. The drawback of this approach is that it requires substantially more samples to produce tight bounds. This implies longer acquisition times, which may be experimentally challenging due to inevitable temperature fluctuations.

We will use the methods described in this section to determine the minimum $q_\mathrm{rel}$-parameter $q^\mathrm{min}_\mathrm{rel}$ as a function of the optical power injected by Eve in Sec.~\ref{sec:max_degree_il}. However, we first discuss a technical aspect related to the influence of the ML polarization on the degree of injection locking in the following section. 

\subsection{Polarization-dependency of injection locking} \label{sec:polarization}

\begin{figure*}[t]
    \begin{minipage}{\textwidth}
        \centering
        \includegraphics{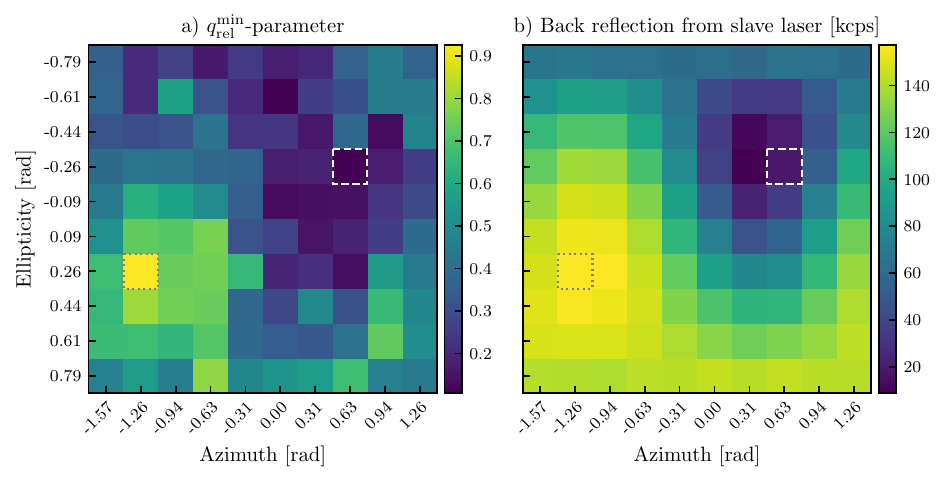} 
        \caption{$q_\mathrm{rel}^\mathrm{min}$-parameter versus polarization state of the light injected into the slave laser. The dashed (dotted) lines mark the polarization states with minimum (maximum) $q_\mathrm{rel}^\mathrm{min}$-parameter. The injected optical power was $-55.3$~dBm. a) Heatmap of the $q_\mathrm{rel}^\mathrm{min}$-parameter versus polarization state of the injected light. b) Heatmap of the single-photon detector count rate while measuring the reflected light at the slave laser cavity while the slave laser was turned off versus the polarization state of the injected light.}
        \label{fig:q_param_vs_polar}
    \end{minipage}
\end{figure*}

\noindent In general, the amount of optical power coupling into the DUT laser cavity depends on the polarization of the injected light. Thus, for optimal injection locking, the best polarization state of the injected light has to be found. With the light sources used in this work, we expect that the ML and SL light should match in polarization for the best coupling. However, our method does not rely on this assumption. To find the best polarization state (and thus replicate Eve's best attack strategy), we measure the polarization of the ML light at the input of the SL (see Fig.~\ref{fig:characterization_setup}) and simultaneously acquire the heterodyne detection signal. 

This procedure is repeated for a series of polarization states to obtain a scan of the Poincaré sphere. For each point on the sphere, we determine the corresponding $q_\mathrm{rel}^\mathrm{min}$-parameter following Sec.~\ref{sec:determine_q_param}. The outcome is shown in Fig.~\ref{fig:q_param_vs_polar} a). The polarization state associated to the smallest $q_\mathrm{rel}^\mathrm{min}$-parameter can then be used for subsequent measurements of the $q_\mathrm{rel}^\mathrm{min}$-parameter e.g. when modifying the attenuation between ML and SL using the VOA\footnote{Importantly, it has to be ensured that modifying the VOA attenuation does not affect the polarization state of the ML. Otherwise, the best polarization state must be determined for each attenuation individually, following the methods described in this section.}. While this measurement provides the best polarization state for the injected light with minimal assumptions on the DUT, the process may take a considerable amount of time. As an example, the scan, whose results are depicted in Fig.~\ref{fig:q_param_vs_polar} a), took around $\SI{5}{\hour}$ to complete. If possible, it is preferable to use a faster method to find the best polarization state.

Another method for finding the optimal polarization is to measure the power of the ML light reflected at the cavity of the SL while it is not biased nor modulated. The intuition behind this method is that the active material is absorptive when the DFB laser is off. This way, we can determine how well the ML couples into the SL laser cavity. To confirm this, we measure the reflected light while determining the $q_\mathrm{rel}^\mathrm{min}$-parameter as a function of the polarization state. Since reflections at the cavity are low, we used a single-photon detector (SPD) for this task (see Fig.~\ref{fig:characterization_setup}). The result of this measurement is depicted in Fig.~\ref{fig:q_param_vs_polar} b) and indicates that for the DFB lasers used in this work, the power of the reflected light is indeed lowest when the polarization state of the ML yields the smallest $q_\mathrm{rel}^\mathrm{min}$-parameter, up to experimental uncertainty. Based on this result, we may now minimize the power of the reflected light to find the best polarization state of the injected light. This method is preferable if the DUT is a DFB laser, since it only takes in the order of minutes to find the optimal polarization, compared to hours for the first approach. 

For the DFB lasers used in this work, both methods are equivalent and yield comparable results, cf. Fig.~\ref{fig:q_param_vs_polar}. However, if the DUT is a more complex optical system, it has to be verified that the above methods can be used interchangeably. For example, if the DUT exhibits polarization dependent losses outside the laser cavity, the second method is not applicable in a straightforward manner, but the first method still is. To summarize, the first polarization optimization method is general and applicable for a broad range of DUTs but takes considerably longer to complete than the second method, i.e. analyzing the back-reflections, which is however only limited to DUTs where the correlation between back-reflections and the $q_\mathrm{rel}^\mathrm{min}$-parameter is known. For all measurements discussed in the remainder of this work, we use the second method.

\begin{figure*}[t]
    \begin{minipage}{\textwidth}
        \centering
        \includegraphics{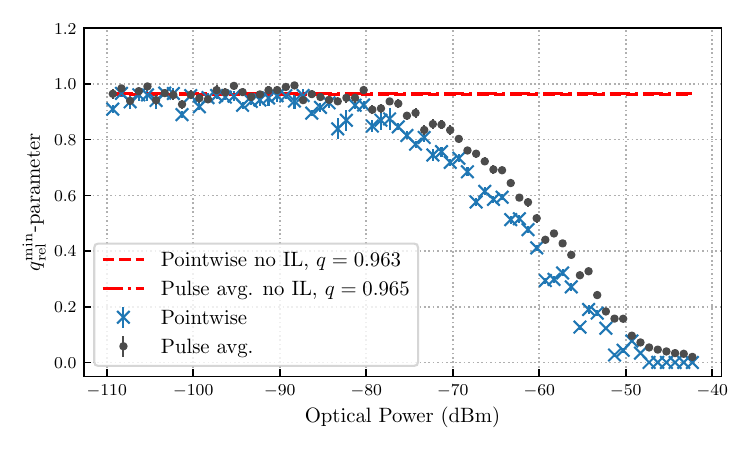} 
        \caption{Degree of phase de-randomization, quantified by the $q_\mathrm{rel}^\mathrm{min}$-parameter defined in Eq.~\eqref{eq:def_q_rel_min}, as a function of the optical power injected into the slave laser. IL: injection locking.}
        \label{fig:q_parameter_over_attenuation}
    \end{minipage}
\end{figure*}

\subsection{Maximum degree of injection locking}
\label{sec:max_degree_il}
\noindent After determining the optimal polarization state of the ML following previous section, we determine the minimum $q_\mathrm{rel}$-parameter $q_\mathrm{rel}^\mathrm{min}$, cf. Sec.~\ref{sec:determine_q_param}, as a function of the injected optical power, for the optimal polarization. In order to achieve this, we incrementally increase the attenuation at the VOA, cf. Fig.~\ref{fig:characterization_setup}, to simulate Alice's components (e.g. fix attenuators, band pass filters, isolators, etc...). The result is depicted in Fig.~\ref{fig:q_parameter_over_attenuation}. For completeness, we also plot the $q_\mathrm{rel}^\mathrm{min}$-parameter resulting from integrating the homodyne detection signal, cf. Eqs.~\eqref{eq:I_0_def} and \eqref{eq:I_pi_2_def}, over the pulse duration. We can see that, for the DFB lasers used in this work, the ML does not significantly influence the phase randomization of the SL if the injected optical power is below approx. $-90$\,dBm.

The error bars for the $q_\mathrm{rel}^\mathrm{min}$-parameters depicted in Fig.~\ref{fig:q_parameter_over_attenuation} result from applying the bootstrapping method to the the phase values obtained in the experiment \cite{Huet2004}. We resampled the data 50 times. An additional source of error which is not accounted for in Fig.~\ref{fig:q_parameter_over_attenuation} is a drift in the difference in central frequencies of the ML and SL on a scale which cannot be resolved by the used OSA, which has a resolution of \SI{0.05}{\nm}.

When considering attacks based on light injection, the maximum optical power Eve can inject is commonly defined by the laser-induced damage threshold (LIDT) of the optical fiber \cite{Makarov2023}. Taking into account the total attenuation of Alice's components, this upper-bounds the optical power reaching Alice's laser. As an example, considering a standard telecommunication optical fiber with laser-induced damage threshold of $100$\,W \cite{Makarov2023}, Alice requires at least about $140$\,dB of attenuation in order to ensure Eve's injection-locking attack does not significantly affect the phase randomization of her laser, following Fig.~\ref{fig:q_parameter_over_attenuation}. 

As discussed in Sec.~\ref{sec:setup}, we recall that an important aspect of the proposed method is its reliance on the ML's ability to simulate Eve's optimal attack, which fundamentally limits the characterization. Without further assumptions and with current security proofs necessitating at least a partial characterization of the phase distribution, developing a method to characterize the maximum degree of injection locking without dependency on the ML used is challenging and requires further experimental and theoretical investigations.

We note that the methods described in this work can be used to evaluate QKD systems against injection-locking attacks through means of black-box testing. In this case, we replace the slave laser in Fig.~\ref{fig:characterization_setup} by the QKD system under test and inject light with the optimal polarization state, which is determined following the discussion in Sec.~\ref{sec:polarization}. The QKD system should be operated such as to facilitate the evaluation and simulate the best possible scenario for Eve. For example, we turn off any modulation of components other than the laser, e.g. intensity modulators and VOAs, to ensure that only the gain-switching of the laser shows a time-dependency, and set any component affecting the attenuation of Alice's setup to the highest transmission that can arise during operation. We then verify that for the maximum chosen optical power injected, the resulting $q_\mathrm{rel}^\mathrm{min}$-parameter is greater than the $q$-parameter and/or $q_\mathrm{rel}^\mathrm{min}$-parameter provided by the implementer. Indeed, by definition, $q_\mathrm{rel}^\mathrm{min} \geq q$, cf. Sec.~\ref{sec:pulse_phase}. 

\section{Conclusion}\label{sec:conclusion}
\noindent We have presented a general method to determine the maximum degree of phase de-randomization Eve can induce from an injection-locking attack, quantified by the $q_\mathrm{rel}$-parameter. This approach is source-agnostic, making it applicable to any QKD system that uses optical pulses, e.g. those implementing the decoy-state BB84 protocol \cite{Lim14, Rusca18}. An important feature of our method is that its effectiveness in characterizing the maximum degree of injection locking is directly given by how well the master laser can emulate Eve's optimal attack strategy. 

Using the DFB lasers in this work, we observed that a minimum attenuation of approx. $140\,$dB is required to largely protect Alice from an injection-locking attack. However, it is crucial to perform system-specific characterizations for each QKD implementation, as the vulnerability to such attacks can drastically vary depending on the light source employed. We believe our methods remain valid for a broad range of master lasers and DUTs. The methods presented in this work can be used for black-box testing, serving as a valuable tool for evaluating QKD systems against injection-locking attacks. 

Finally, we emphasize that while evaluating QKD systems against side-channels attacks is certainly an important step toward certifiable (and certified) QKD systems, device imperfections (e.g. injection locking) must be thoroughly characterized and/or rigorously taken into account in the security proof. Following the discussions in this work, the next step is to determine the maximum degree of injection locking Eve can induce \textit{independently} of the devices used to perform the characterization.
\section*{Acknowledgements} \label{sec:acknowledgements}
\noindent We thank Hugo Zbinden and Margarida Pereira for fruitful discussions. The authors are subcontractors in the EC funded project, Nostradamus, TOPIC ID: CNECT/2023/OP/0032. It is the goal of Nostradamus to describe the blueprint for a Testing \& Validation Infrastructure to enable the evaluation and certification of QKD devices and related technologies, as well as to implement and operate a prototypical testbed facility to offer initial evaluation services which are mandatory for the accreditation from a European security authority. The authors would like to thank the whole project team for the support and valuable exchange. Views and opinions expressed are those of the authors only and do not necessarily reflect those of the European Union or the European Commission. Neither the European Union nor the granting authority can be held responsible for them. This work was supported by the Galician Regional Government (consolidation of Research Units: AtlantTIC), MICIN with funding from the European Union NextGenerationEU (PRTR-C17.I1) and the Galician Regional Government with own funding through the “Planes Complementarios de I+D+I con las Comunidades Autónomas” in Quantum Communication and
the European Union’s Horizon Europe Framework Programme under the project “Quantum Security Networks Partnership” (QSNP, grant agreement No 101114043).
\section*{Author contributions}
\noindent D.R. and N.W. supervised the research. J.W. and F.G. performed the research. D.R. assisted and supported the research. A.B. provided part of the setup and assisted in building the setup. F.G. and J.W. analyzed the data. J.W. and F.G. wrote the manuscript. All authors participated in discussions and reviewed the manuscript.

\appendix

\section{Diagonal form of the signal density matrix} \label{ap:diagonal_form}
\noindent The density matrix representation from Eq.~\eqref{eq:density_matrix_fock} follows from Eq.~\eqref{eq:density_matrix_coherent} if the phases are uniformly distributed, i.e. $f(\theta) = 1/(2\pi)$. To see this, we recall that a coherent state of one mode of the electromagnetic field can be written as a superposition of Fock states \cite{Glauber1963}
\begin{equation}
    \ket{\alpha} = \sum_{n=0}^\infty e^{-|\alpha|^2/ 2} \frac{\alpha^n}{\sqrt{n!}}\ket{n}.
    \label{eq:def_coherent_state}
\end{equation}
We can now write $\ket{\alpha}=\ket{\sqrt{\mu}e^{i\theta}}$, where $\mu$ is the mean photon number and $\theta$ denotes the phase of the state. Plugging this expression into Eq.~\eqref{eq:density_matrix_coherent}, where $f(\theta)$ describes a uniform distribution, yields
\begin{equation}
    \rho_\mu = \int_0^{2\pi} \frac{d\theta}{2\pi} \sum_{n=0}^\infty \sum_{m=0}^\infty e^{-\mu} e^{i(n\theta - m\theta)} \frac{\sqrt{\mu}^n\sqrt{\mu}^m}{\sqrt{n!m!}}\ket{n}\bra{m}.
\end{equation}
Identifying the representation of the Kronecker delta $\delta_{nm} = (2\pi)^{-1} \int_0^{2\pi} d\theta e^{i(n\theta - m\theta)}$ results in 
\begin{equation}
    \rho_\mu = \sum_n e^{-\mu}\frac{\mu^n}{n!}\ket{n}\bra{n},
\end{equation}
which corresponds to a coherent superposition of Fock states, i.e. Eq.~\eqref{eq:density_matrix_fock}, and shows that the density matrix is diagonal in the Fock basis.

\section{List of components}
\label{ap:list_equipment}
\noindent We list all the components used to perform the characterization, i.e. the components depicted in Fig.~\ref{fig:characterization_setup}:
\begin{itemize}
    \item Master laser: Goch and Housego AA0701-193414-010-SM900-FCA-50 1550nm
    \item Slave laser: LP-PD LP-ML1001A-55-FA
    \item Optical spectrum analyzer: Yokogawa Spectrum Analyzer AQ6375E-10-L1-F/FC/RFC
    \item Polarimeter: Thorlabs PAX1000IR2/M
    \item Balanced detectors: Thorlabs PDB480C-AC
    \item Manual polarization controller: Thorlabs FPC560
    \item Motorized polarization controller: Thorabs MPC320
    \item Programmable temperature controller: Stanford research systems PTC10
    \item Single-photon avalanche diode: IDQube-NIR-FR-MMF-LN
    \item Arbitrary waveform generator: Tektronix AFG31102
    \item Variable optical attenuator: EXFO FVA-600
    \item Mixed signal oscilloscope: Tektronix MSO64B 6-BW-4000 Installed Option, 4 GHz Bandwidth
    \item 90$\degree$ optical hybrid: iXblue COH24
    \item Laser diode controller: Newport LDX-3412-240V
    \item Circulator: Thorlabs 6015-3-APC
\end{itemize}

\section{Homodyne detection signal}\label{ap:homodyne_detection_signal}
\noindent Assume generic electric fields $\boldsymbol{E}_\text{S}(t)$ and $\boldsymbol{E}_\text{LO}(t)$ at a point in space for the signal and local oscillator. Following their interference, the measured intensities at both arms of the interferometer are
\begin{align}
    I_1(t) &= \frac{1}{2}\left|\boldsymbol{E}_\text{S}(t) + \boldsymbol{E}_\text{LO}(t)\right|^2\,,\\
    I_2(t) &= \frac{1}{2}\left|\boldsymbol{E}_\text{S}(t) + e^{i\pi}\boldsymbol{E}_\text{LO}(t)\right|^2\,,
\end{align}
and the resulting homodyne detection signal is $I_0(t) = I_1(t) - I_2(t)$. Here, we assume perfectly balanced detectors. Now, if a $90\degree$ optical hybrid is used as in Fig.~\ref{fig:characterization_setup}, then the electric fields in the arms are 
\begin{align}
    I_3(t) &= \frac{1}{2}\left|e^{-i\frac{\pi}{2}}\boldsymbol{E}_\text{S}(t) + \boldsymbol{E}_\text{LO}(t)\right|^2 \,,\\
    I_4(t) &= \frac{1}{2}\left|\boldsymbol{E}_\text{S}(t) + e^{-i\frac{\pi}{2}}e^{i\pi}\boldsymbol{E}_\text{LO}(t)\right|^2 \,,
\end{align}
which leads to a homodyne detection signal $I_{\pi/2}(t) = I_3(t) - I_4(t)$.

\section{Time window for $q_\mathrm{rel}$-parameter determination}
\label{app:time_window_argumentation}
\noindent The sum of the squared residuals is defined as
\begin{align}
S^2 &= \sum_{n=1}^N 
    \Big(f_w\left(\Delta\theta_n,\mu_{\mathrm{opt}},\sigma_{\mathrm{opt}},\gamma_{\mathrm{opt}}\right) - c_n
\Big)^2\,,
\end{align}
where $\Delta \theta_n$ is $\Delta\theta^{(n)}{(\tau_n)}$ evaluated at the time which minimizes $q_\text{rel}^\text{min}$, $c_n$ is the amount of counts in the $n$-th histogram bin (an example is shown in Fig.~\ref{fig:histogram}) and $\mu_{\mathrm{opt}}$, $\sigma_{\mathrm{opt}}$ and $\gamma_{\mathrm{opt}}$ are the optimal parameters found by the non-linear regression fit when determining $q_\text{rel}^\text{min}$.

\begin{figure}[h]
    \centering
    \includegraphics{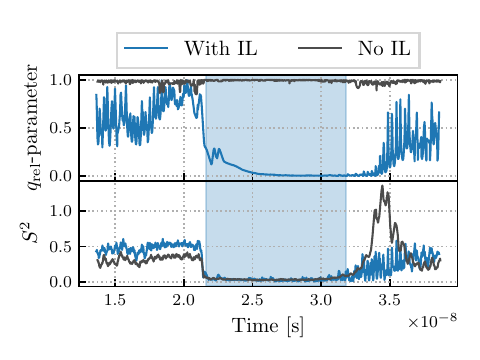}
    \caption{$q_\mathrm{rel}$-parameter and sum of squared residuals $S^2$. The time window for the determination of the $q_\mathrm{rel}$-parameter is shaded in light blue.}
    \label{fig:residuals}
\end{figure}

To determine the minimum $q_\mathrm{rel}$-parameter $q_\mathrm{rel}^\mathrm{min}$, cf. Eq.~\eqref{eq:def_q_rel_min}, we selected the time window highlighted in light blue in Fig.~\ref{fig:exemplary_pulse}. This selection ensures that our model, described in Eq.~\eqref{eq:pdf_model}, agrees with the observed data and produces reliable results. The goodness of the fit is quantified by the sum of the squared residuals $S^2$ and its behavior with respect to the time $\tau_n$ is shown in Fig.~\ref{fig:residuals}, together with the $q_\mathrm{rel}$-parameter.

\bibliographystyle{unsrt}
\bibliography{lit}{}

\end{document}